\documentclass[12pt]{iopart}
\usepackage[usenames,dvipsnames]{color}
\usepackage{graphicx}
\usepackage{iopams}

\begin{document}
%
\title[]{Coexistence of absolute negative mobility and anomalous diffusion}
\author{J. Spiechowicz$^1$, P. H\"anggi$^{2,3}$, J. {\L}uczka$^1$}
\address{$^1$ Institute of Physics and Silesian Center for Education and Interdisciplinary Research, University of Silesia, 41-500 Chorz{\'o}w, Poland\\
$^2$ Institute of Physics, University of Augsburg, D-86135 Augsburg, Germany\\
$^3$ Nanosystems Initiative Munich, Schellingstr. 4, D-80799 Mu\"nchen, Germany}
\ead{j.spiechowicz@gmail.com}
\begin{abstract}
Using extensive numerical studies we  demonstrate that absolute negative mobility of  a Brownian particle (i.e. the net motion into the direction {\it opposite} to a constant biasing force acting around zero bias) does coexist with anomalous diffusion. The latter is characterized in terms of a nonlinear scaling with time of the mean-square deviation of the particle position. Such anomalous diffusion covers "coherent" motion (i.e. the position dynamics $x(t)$ approaches in evolving time a {\it constant} dispersion), ballistic diffusion, subdiffusion, superdiffusion and hyperdiffusion. In providing evidence for this coexistence  we consider a paradigmatic model of an inertial Brownian particle moving in a one-dimensional symmetric periodic potential being driven by both an unbiased time-periodic force and a constant bias. This very setup allows for various sorts of different physical realizations.
\end{abstract}
\maketitle
\section{Introduction}
Phenomena of anomalous transport often appear to be paradoxical, non-intuitive and contrary to our everyday observations. Such behaviour has been identified in a rich variety of systems originating from all branches of natural sciences including, most of all, physics, chemistry and physical biology \cite{klages2008}. Generally, there exist two types of transport phenomena: passive diffusion and active transport. The former class is undirected as it is driven by thermal equilibrium fluctuations due to random collisions with  surrounding fluid molecules \cite{hanggi2005}. On the other hand, active transport is directional but at the cost of supplied energy from an external source in order to take the system far from equilibrium \cite{lutz2012,lutz2015,bechinger2016}. The latter may be also provided by a non-passive environment as active matter systems typically possess a property which allow them to extract energy from their surroundings \cite{bechinger2016}. These two classes of transport phenomena usually coexist since directed motion is commonly accompanied by erratic diffusion.

The phenomenon of absolute negative mobility (ANM)  \cite{eichhorn2002a,eichhorn2002b,machura2007,nagel2008,kostur2008,cecconi,ai,mukamel,chen,slapik2019} may be regarded as a form of anomalous directed transport. In this case, transported particles move in a direction opposite to the net acting force around zero bias. On the other hand, anomalous diffusion (AD) is characterized  as the time-nonlinear scaling of the spread of particle trajectories around its mean path \cite{sokolov2005,metzler2014,oliv}. These two anomalies inspired the entire research fields which have been independently explored extensively over recent decades.
In this paper, we would like to address the problem whether ANM and AD can coexist and thereby be simultaneously detected in physical systems. This being the case we want to unite these two forms of anomalous transport phenomena to investigate their interrelation(s) and draw attention to this challenging objective. In doing so we consider an archetype of diffusion processes, an inertial Brownian particle dwelling in a periodic potential \cite{risken1996}. The physical phase space for this system is rather complex, as described 
by a set $\{x, v\}$ including all positions $x$ and velocities $v$ of the Brownian particle. 
ANM takes place in the velocity sub-space and AD occurs in the position sub-space. We shall search the phase space $\{x, v\}$ to find sets $\{\mbox{AD}, \mbox{ANM}\}$ immersed in it.

This work is organized as follows. In section 2 we introduce the model as well as the corresponding quantifiers describing both  directed transport and anomalous diffusion. The next section contains the detailed discussion on the problem of coexistence of ANM and AD. The last section is devoted to our conclusions.

\section{Description of the model}

The conceptually simplest model, which exhibits both ANM and AD,  is a classical Brownian particle of mass $M$ travelling in a one-dimensional, spatially  periodic potential $U(x)$ and subjected to both, unbiased time-periodic driving $F_d(t)$ in combination with  a constant, biasing force $F$. This setup has been used in the literature before  in investigating various phenomena including, for example the Brownian motor or the stochastic ratchet effect \cite{hanggi2009,cubero2016}, ANM \cite{machura2007,nagel2008}, a non-monotonic temperature dependence of a diffusion coefficient in normal diffusion
\cite{marchenko2012,spiechowicz2016njp,spiechowicz2017chaos} or also
transient time-dependent AD \cite{spiechowicz2016scirep}, to name but a few. This model is described  by the Langevin equation of the form
\begin{equation}
	\label{model}
	M\ddot{x} + \Gamma\dot{x} = -U'(x) + F_d(t) + F + \sqrt{2\Gamma k_B T}\,\xi(t),
\end{equation}
where $\Gamma$ denotes the friction coefficient, the dot and the prime denotes a differentiation with respect to  time and the particle coordinate $x$, respectively. The potential $U(x)=U(x+L)$ is assumed to be \emph{symmetric} of spatial period $L$ and the barrier height is $2 \Delta U$, reading
\begin{equation}
	\label{potential}
	U(x) = -\Delta U\sin{\left( \frac{2\pi}{L}x \right)}.
\end{equation}
The time periodic force of  amplitude $A$ and angular frequency $\Omega$  is $F_d(t) = A\cos{(\Omega t)}$.
Thermal equilibrium fluctuations due to interaction of the particle with its environment of temperature $T$ are modelled  as  $\delta$-correlated Gaussian white noise of vanishing mean
\begin{equation}
	\langle \xi(t) \rangle = 0, \quad \langle \xi(t)\xi(s) \rangle = \delta(t - s).
\end{equation}
The  noise intensity  $2\Gamma k_B T$ in equation (\ref{model}) follows from  the fluctuation-dissipation theorem \cite{kubo,thomas}, where $k_B$ is the Boltzmann constant.
It thereby ensures the canonical Gibbs state in thermal equilibrium for $A = 0$ and $F = 0$. If $A \neq 0$, then the harmonic force $F_d(t)$ drives the system far away from an equilibrium state and the setup serves as a model of nonequilibrium setting. The complexity of the stochastic dynamics is routed in the wide parameter space, involving the values for $M, A, \Omega, F, \Gamma, T$ in presence of nonequilibrium driving of the underlying inertial stochastic dynamics with acting thermal fluctuations of varying strength.

There are several physical systems that can be modelled by  a dynamics as given by (\ref{model}). Pertinent examples which come to mind are the semiclassical dynamics of the phase difference across a Josephson junction and its variations including e.g. the SQUIDs \cite{kautz1996, blackburn2016, spiechowicz2014prb} as well as the dynamics exhibited by  cold atoms dwelling in optical lattices \cite{renzoni} or the transport of ions in superionic conductors \cite{fulde1975,dieterich1980}.

Next, we  rewrite  the inertial Langevin  dynamics in its dimensionless form. To this aim, we use the following scales as  characteristic units of length and time
\begin{equation}
	\label{scaling}
	\hat{x} =2\pi  \frac{x}{L}, \quad \hat{t} = \frac{t}{\tau_0}, \quad \tau_0 = \frac{L}{2\pi}\sqrt{\frac{M}{\Delta U}}.
\end{equation}
Under such a procedure  equation (\ref{model}) is transformed into the form
\begin{equation}
	\label{dimless-model}
	\ddot{\hat{x}} + \gamma\dot{\hat{x}} = -\hat{U}'(\hat{x}) + a\cos{(\omega \hat{t})} + f + \sqrt{2\gamma Q} \hat{\xi}(\hat{t}).
\end{equation}
In this scaling the dimensionless mass $M$ of the particle assumes unity, i.e. $m = 1$. The remaining five dimensionless parameters  read
\begin{eqnarray}
	\gamma = \frac{\tau_0}{\tau_1}, \quad a = \frac{1}{2\pi}\frac{L}{\Delta U}\, A, \quad f = \frac{1}{2\pi}\frac{L}{\Delta U}\, F, \quad Q = \frac{k_B T}{\Delta U}, \quad \omega = \tau_0 \Omega,
\end{eqnarray}
where the second characteristic time scale is $\tau_1 = M/\Gamma$. This second timescale has the physical interpretation of the relaxation time for the velocity of the free Brownian particle. On the other hand, the time scale $\tau_0$ can be viewed as the characteristic period of oscillations inside the potential wells.

The dimensionless potential
 $\hat{U}(\hat{x}) = U((L/2\pi)\hat{x})/\Delta U$ possesses the period $L=2\pi$. The re-scaled thermal noise reads explicitly \mbox{$\hat{\xi}(\hat{t}) = (L/2\pi \Delta U)\xi(t) = (L/2\pi \Delta U)\xi(\tau_0\hat{t})$} and assumes the same statistical properties as $\xi(t)$; i.e., $\langle \hat{\xi}(\hat{t}) \rangle = 0$ and \mbox{$\langle \hat{\xi}(\hat{t})\hat{\xi}(\hat{s}) \rangle = \delta(\hat{t} - \hat{s})$}.
In the following we shall stick to these  dimensionless variables throughout. In order to simplify the notation further, we shall omit the $\wedge$-notation in the above equation (\ref{dimless-model}).

\subsection{Quantifiers for ANM and AD}

ANM is characterized by the long-time averaged velocity of the Brownian particle. Due to the presence of both the external time-periodic driving $a\cos{(\omega t)}$ and the dissipation $\gamma\dot{x}$ the particle velocity $\dot{x}(t) = v(t)$ approaches a unique nonequilibrium asymptotic state being characterized by a temporally periodic probability density, namely
\begin{equation}
	P_{as}(v, t) = P_{as}(v, t + \mathsf{T}),
\end{equation}
where
\begin{equation}
	P_{as}(v, t) = \int_{-\infty}^{+\infty} dx \,P_{as}(x,v,t)
\end{equation}
and
\begin{equation}
	P_{as}(x, v, t) = \lim_{t \gg 1} P(x, v, t).
\end{equation}
Using Floquet theory, this function assumes the same period $\mathsf{T} = 2\pi/\omega$ as the external driving \cite{jung1993}. For this reason the average value of the instantaneous particle velocity for the asymptotic long time regime $\langle v(t) \rangle_{as}$ can be represented as 
\begin{equation}
\label{asym}
	\langle v(t) \rangle_{as} = \int_{-\infty}^{+\infty} dv \, v \, P_{as}(v, t).
\end{equation}
The most important quantifier is the time-independent directed velocity  $\langle \mathbf{v} \rangle$ while the contribution of higher time-periodic harmonics {\it vanishes} after time-average over the fundamental period $\mathsf{T} = 2\pi/\omega$, leading to 
\begin{equation}
	\langle \mathbf{v} \rangle = \frac{1}{\mathsf{T}}\int_t^{t + \mathsf{T}} ds \, \langle v(s) \rangle_{as} = \int_{-\infty}^{+\infty} dv\, v \,P_{as}(v),
\end{equation}
where we introduced the time averaged asymptotic velocity distribution
\begin{equation}
	P_{as}(v) = \frac{1}{\mathsf{T}}\int_t^{t + \mathsf{T}} ds \, P_{as}(v,s).
\end{equation}
On the other hand, since we are interested not only in the asymptotic state but also in the full time dynamics it is useful to consider the period averaged velocity
\begin{equation}
	\langle \mathbf{v}(t) \rangle = \frac{1}{\mathsf{T}} \int_{t}^{t + \mathsf{T}} ds \,\langle v(s) \rangle,
\end{equation}
which relates to the directed velocity $\langle \mathbf{v} \rangle$ in the following way
\begin{equation}
	\langle \mathbf{v} \rangle = \langle \mathbf{v}(t) \rangle_{as} = \lim_{t \gg 1} \,\langle \mathbf{v}(t) \rangle,
\end{equation}
where $\langle \cdot \rangle$ indicates averaging with respect to the probability measure $P(x,v,t)$, i.e., over all thermal noise realizations as well as over initial conditions for the position $x(0)$ and the velocity $v(0)$.
%
%

Diffusion is characterized by the  mean-square deviation (variance) of the particle position $x(t)$, namely,
\begin{equation}
	\label{msd}
	\langle \Delta x^2(t) \rangle = \langle \left[x(t) - \langle x(t) \rangle \right]^2 \rangle = \langle x^2(t) \rangle - \langle x(t) \rangle^2.
\end{equation}
If the long time behaviour assumes a linear function of time we term it as normal diffusion. Any departure from strict linearity  at asymptotic times  qualifies as a process exhibiting anomalous diffusion. For example, if the variance  becomes an increasing function of elapsing time and grows according to the power law \cite{metzler2014}
\begin{equation}
	\label{alpha}
	\langle \Delta x^2(t) \rangle \sim t^{\alpha}
\end{equation}
then normal diffusion is observed for $\alpha = 1$. On the other hand, if $0 < \alpha < 1$ is termed  subdiffusion while for $\alpha > 1$ we classify this behaviour as superdiffusion. Let us define also
the time-dependent "diffusion coefficient" as
\begin{equation}
	\label{diffusioncoefficient}
	D(t) = \frac{\langle \Delta x^2(t) \rangle}{2t}.
\end{equation}
Note that the case of time-decreasing $D(t)$ at long times corresponds to subdiffusion whereas superdiffusion occurs when $D(t)$ increases when $t$ grows at large evolving times. For $D(t) = const.$ asymptotically we deal with normal diffusion. We stress that only in the asymptotic long time regime with the exponent $\alpha$ approaching unity we find a properly defined, finite diffusion coefficient $D$, i.e.,
\begin{equation}
	D = \lim_{t \to \infty} D(t) < \infty.
\end{equation}
If the diffusion process is anomalous then $D(t)$ either diverges to infinity (superdiffusion) or converges to zero (subdiffusion) when $t\to\infty$.  The power law grow for spread of particle trajectories  can be inverted to obtain the instantaneous exponent
\begin{equation}
	\alpha(t) = t \,\frac{d \ln{\langle \Delta x^2(t) \rangle}}{dt}
\end{equation}
which helps to identify the diffusion anomalies during the system evolution. However, due to the temporal periodicity of the corresponding probability distribution, see Eq. (7)-(9), the contribution of higher harmonics to this quantity can be smoothed upon an average over  the fundamental external driving  period, i.e.,
\begin{equation}
	\boldsymbol{\alpha}(t) = \frac{1}{\mathsf{T}} \int_{t}^{t + \mathsf{T}} ds \,\alpha(s).
\end{equation}
This  exponent in turn tends to a constant value in the asymptotic long time regime, reading
\begin{equation}
	\boldsymbol{\alpha} = \lim_{t \gg 1} \,\boldsymbol{\alpha}(t).
\end{equation}

\section{Coexistence of ANM and AD}

The probability density $P(x, v, t)$ for the particle coordinate $x$ and its velocity $v=\dot x$ obeys a Fokker-Planck equation corresponding to the Langevin equation (\ref{dimless-model}). It is a parabolic partial differential equations with a time-periodic drift coefficient in phase space of position and velocity. Given the complexity with a nonlinear periodic potential $U(x)$  together with a five-dimensional parameter space analytic time-dependent  solutions become unattainable. For this reason, in order to systematically analyse the emerging  rich variety of possible transport behaviours we carried out comprehensive numerical simulations. Technical details of the latter are deferred to the Appendix.

From the underlying symmetry of the Langevin equation (\ref{dimless-model}) it follows that the transformation $f \to  -f$ implies $\langle \mathbf{v} \rangle \to -\langle \mathbf{v} \rangle$. In other words, the average velocity $\langle \mathbf{v} \rangle(f)$  is an odd function of  the constant force $f$:
$\langle \mathbf{v} \rangle(-f) =  -\langle \mathbf{v} \rangle (f)$. Consequently,  we limit our consideration to the case $f > 0$. Generally, $\langle \mathbf{v} \rangle$ assumes a  non-linear function of $f$ and it is expected that $\langle \mathbf{v} \rangle$ typically  increases for growing $f$. However, in the parameter space one can encounter also regimes in which the particle moves on average in the direction opposite to the applied bias, namely, $\langle \mathbf{v} \rangle < 0$ for $f > 0$ thus exhibiting ANM.
We next define the 'mobility coefficient' $\mu(f)$ by the relation $\langle \mathbf{v} \rangle = \mu(f) f$. In the linear response regime, the velocity $\langle \mathbf{v} \rangle = \mu_0 f$, is a linear function of the force $f$ and $\mu_0$ denotes a mobility coefficient being independent of $f$. In the ANM regime, however, the 'mobility coefficient' assumes paradoxical negative values, i.e. $\mu(f) <0$ for $f>0$.
The key requirement for the occurrence of negative mobility is that the system is driven far from thermal equilibrium into a nonequilibrium state \cite{machura2007}. In our case this condition is satisfied by the presence of the external time-periodic driving $a\cos{(\omega t)}$. It has been shown that there exist two fundamentally different mechanisms responsible for the emergence of negative mobility with our setup: (i) it can be generated by the chaotic dynamics or (ii) be induced by thermal equilibrium fluctuations \cite{machura2007}. The former scenario is rooted in the intricate evolution of the corresponding deterministic system given by the Langevin equation (\ref{dimless-model}) with vanishing diffusion $Q = 0$. The latter dynamics can be cast into a set of three autonomous differential equations of first order and the corresponding  phase space is three-dimensional $\{x, \dot{x}, \theta:=\omega t\}$, being the minimal requirement for which the deterministic system displays chaotic evolution \cite{strogatz2014}. Very recently, even a third mechanism generating the negative mobility effect has been discovered. Accordingly, negative mobility phenomenon may occur as well within deterministic non-chaotic parameter regimes \cite{slapik2018}.

As presently well researched, anomalous diffusion (AD) can emerge in various systems, e.g.  in random potentials and disordered  systems \cite{khoury2011, simon2014, hanes2012}. For space-periodic potentials, AD is a transient effect \cite{spiechowicz2016scirep,hanes2013}. However, the  lifetime of AD can be many, many orders of magnitude longer than characteristic times of the system and turns out to be extraordinarily sensitive to the system parameters such as temperature \cite{spiechowicz2016scirep}.
Moreover, it has been demonstrated  \cite{spiechowicz2016scirep} that the  existence of transient superdiffusion for the driven Brownian particle dwelling in the periodic potential can be related to a mechanism known as breaking of strong ergodicity  of the corresponding deterministic counterpart of the system.   On the other hand, the reason for the transient subdiffusive behaviour was identified as a thermal noise induced dynamical localization in the particle velocity (momentum) subspace \cite{spiechowicz2017scirep}. The latter phenomenon leads to the emergence of  power law tails in a residence time probability distribution for the velocity states of the Brownian particle \cite{spiechowicz2019chaos}.

The goal of this paper is to answer the following objective: Is it possible to observe  simultaneously both instances of anomalous transport, namely ANM and AD? At this point it is worth to note that ANM alone is an exceptional phenomenon because even for the most optimal regimes it occurs only in small sets of the  parameter space. Moreover, AD in our system can either be persistent or manifest itself  only as a transient phenomenon.  Therefore it is necessary to distinguish with care an asymptotic long time (time-averaged asymptotic stationary) state from an intermediate (transient)  regime. We will first focus on the former case.
\begin{figure}[t]
\centering
\includegraphics[width=0.45\linewidth]{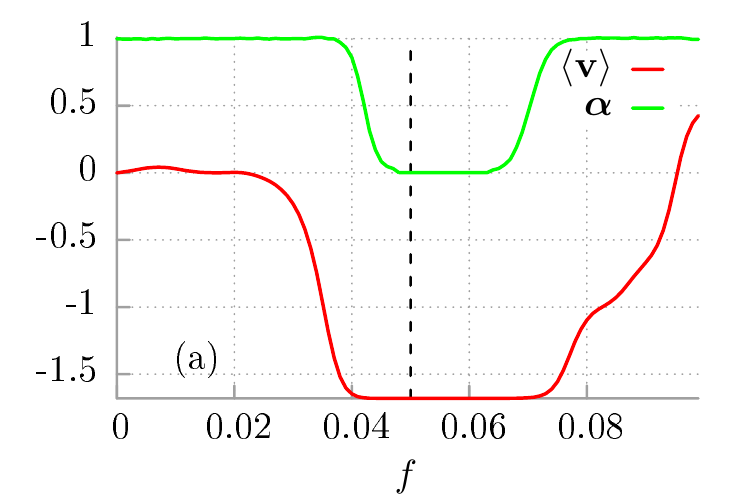}
\includegraphics[width=0.45\linewidth]{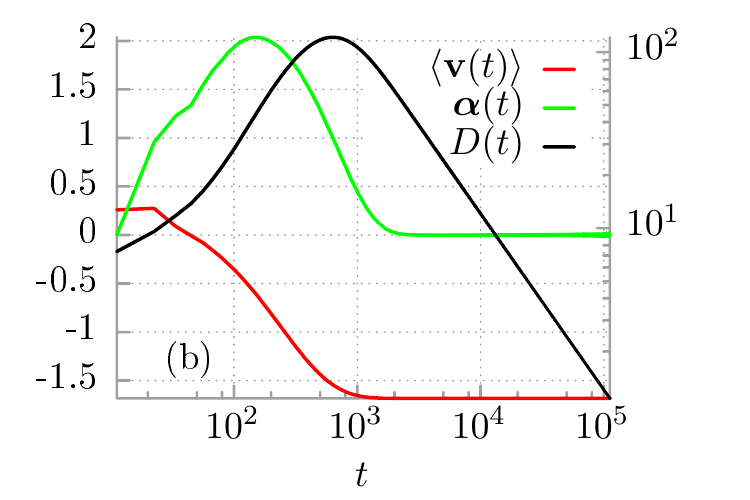}\\
\includegraphics[width=0.45\linewidth]{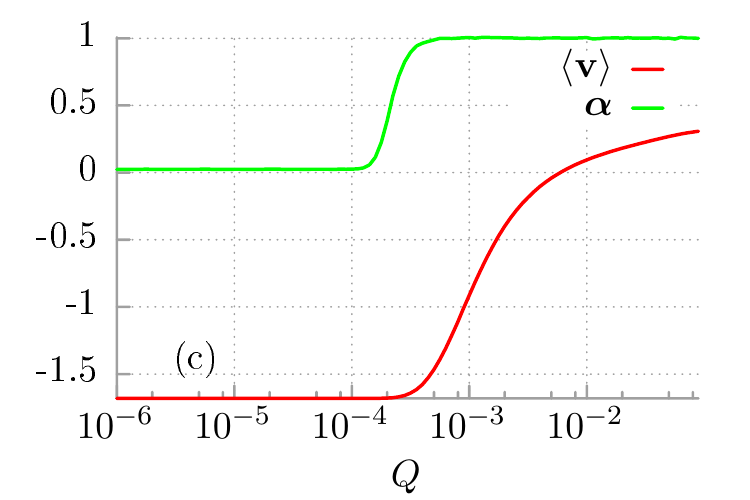}
\includegraphics[width=0.45\linewidth]{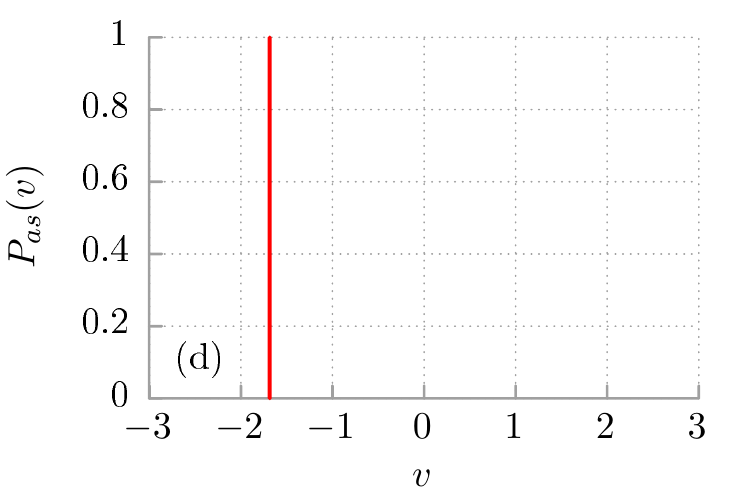}
\caption{Coexistence of absolute negative mobility and transport with constant dispersion. Panel (a): the asymptotic long-time directed velocity $\langle \mathbf{v} \rangle$, Eq. (11), and the power exponent $\boldsymbol{\alpha}$, Eq. (21), are depicted as a function of the external bias $f$. Panel (b): time evolution of the time-averaged velocity over the driving period $\mathsf{T}$, i.e.,
$\langle \mathbf{v}(t) \rangle$, Eq. (13), the power exponent $\boldsymbol{\alpha}(t)$, Eq. (20), (both left axis) as well as the diffusion coefficient $D(t)$, Eq. (17), (right axis). Panel (c): the directed velocity and the power exponent presented versus temperature $Q \propto T$. Panel (d): driving-period averaged asymptotic probability distribution $P_{as}(v)$, Eq. (12), depicted for the deterministic system $Q = 0$. Parameters are: $a = 1.6$, $\omega = 0.561$, $\gamma = 0.174$,  $Q = 10^{-4}$. In panels (b)-(d) $f = 0.05$, at which $\boldsymbol{\alpha} = 0.0026$ in panel (a), as indicated by the vertical dashed line. The fundamental period of driving force equals $\mathsf{T} = 11.2$.}
\label{fig1}
\end{figure}
\subsection{Asymptotic regimes}
\subsubsection{Transport with constant dispersion}

The vast majority of asymptotic regimes for which ANM is observed are accompanied by normal diffusion. However, very rarely AD is encountered. We exemplify this situation with Fig. \ref{fig1} where we demonstrate  coexistence of negative mobility and transport with a constant dispersion. In the latter effect, the spread of Brownian particle trajectories asymptotically assumes a constant \cite{lindenberg2007, saikia2009}. This behaviour is formally described  with a vanishing(!) power exponent  $\alpha = 0$.

In panel (a) we depict the directed velocity $\langle \mathbf{v} \rangle$, Eq. (11), and the averaged power exponent $\boldsymbol{\alpha}$, Eq. (21), as a function of the external bias $f$ in the asymptotic long time regime. We observe that there exists a window of positive values of the force for which the directed velocity of the Brownian particle is negative, thus implying negative mobility $\mu(f) < 0$. For nearly the same interval of the bias the power exponent $\boldsymbol{\alpha}$ drops to zero $\boldsymbol{\alpha} = 0$, indicating transport exhibiting at asymptotic times a constant dispersion. It is necessary to confirm whether the observed diffusion anomaly is only present as a transient phenomenon or in fact  persists up to the asymptotic regime. Towards this goal we study the evolution of the driving-period averaged velocity $\langle \mathbf{v}(t) \rangle$ defined in Eq. (13) and the power exponent $\boldsymbol{\alpha}(t)$, see Eq. (20), which is depicted in panel (b) of Fig. \ref{fig1}. We observe that the transient time span takes about $10^3$ dimensionless time units and after that both quantifiers settle down at their asymptotic stationary values. The conclusion is that negative mobility and  anomalous diffusion  transport with a constant dispersion does coexist in the asymptotic regime. In the same panel we additionally depict the time dependent diffusion coefficient $D(t)$ which corroborates this finding.

To answer the question about the origin of these observed phenomena we examine the temperature dependence of the  directed velocity $\langle \mathbf{v} \rangle$, Eq. (11), and the period averaged power exponent $\boldsymbol{\alpha}$, Eq. (21), which we depict in panel (c) for the case $f=0.05$. It is evident that  negative mobility is induced by the deterministic dynamics as this effect survives even in the low temperature regime $Q \to 0$. 

For the considered noiseless counterpart of the system there are three Lyapunov exponents $\lambda_x, \lambda_{\dot{x}}$ and $\lambda_{\theta:=\omega t}$. As the system is dissipative and its phase space volume is contracting during the evolution,  the sum of these three exponents must be negative,
\begin{equation}
	\lambda_x + \lambda_{\dot{x}} + \lambda_{\theta} < 0.
\end{equation}
The exponent $\lambda_{\theta}$ corresponds to the direction parallel to the particle phase space trajectory and it does not contribute to a change of the phase volume occupied by the setup, i.e. $\lambda_{\theta} = 0$. The remaining exponents are negative or zero in the (quasi-periodic) states, whereas in the chaotic regime one exponent is positive indicating divergence of trajectories. Therefore, in order to detect chaotic behaviour of the system it is sufficient to calculate the maximal Lyapunov exponent $\lambda$ and check its positivity. In the discussed case it is zero and hence  negative mobility is induced by the deterministic non-chaotic mechanism \cite{slapik2018}.

This observation in turn facilitates the explanation for the origin of transport with constant dispersion. We can notice in panel (c) that the coherence is maintained for the low-to-moderate temperature regimes and only thermal noise of high intensity can turn the diffusive behaviour back to normal behaviour. It implies that for the deterministic counterpart of the system there exists only one attractor with negative velocity. After a sufficiently long time all trajectories will eventually follow it and consequently the dispersion will not change with time any longer. This observation is confirmed by  panel (d) where for the deterministic system we depict the asymptotic period averaged probability distribution for the velocity $P_{as}(v)$, see Eq. (12). From the dependence of the power exponent $\boldsymbol{\alpha}$ shown in panel (c) we can see that thermal noise of low and moderate intensity cannot perturb the deterministic dynamics significantly and therefore the motion with constant dispersion is still maintained for this temperature interval.
\begin{figure}[t]
\centering
\includegraphics[width=0.45\linewidth]{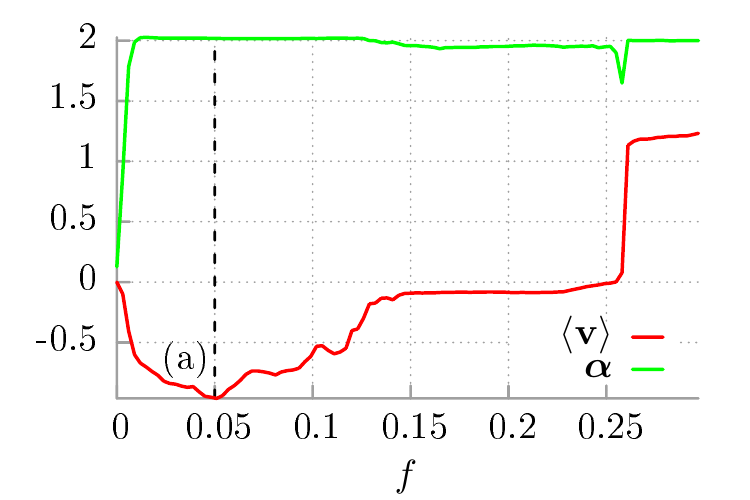}
\includegraphics[width=0.45\linewidth]{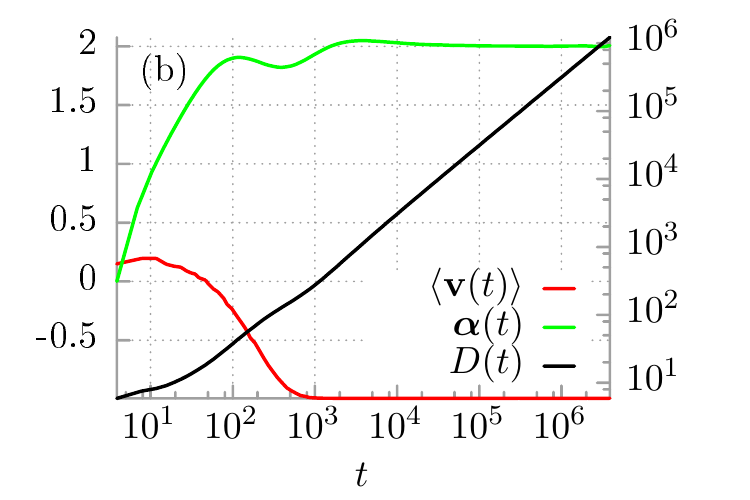}\\
\includegraphics[width=0.45\linewidth]{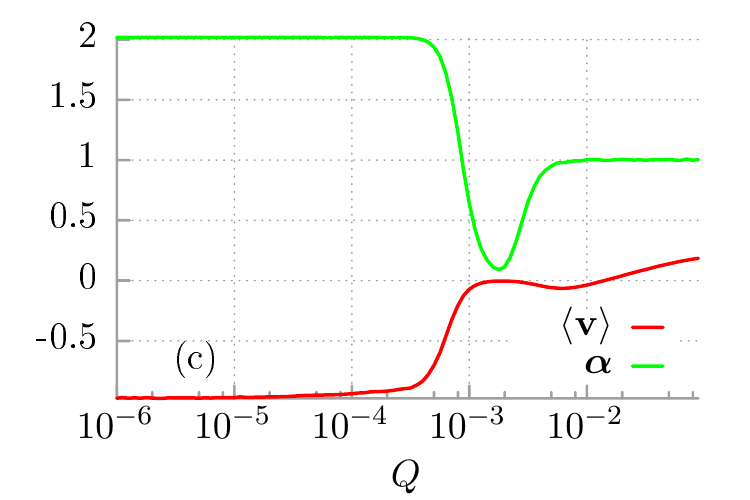}
\includegraphics[width=0.45\linewidth]{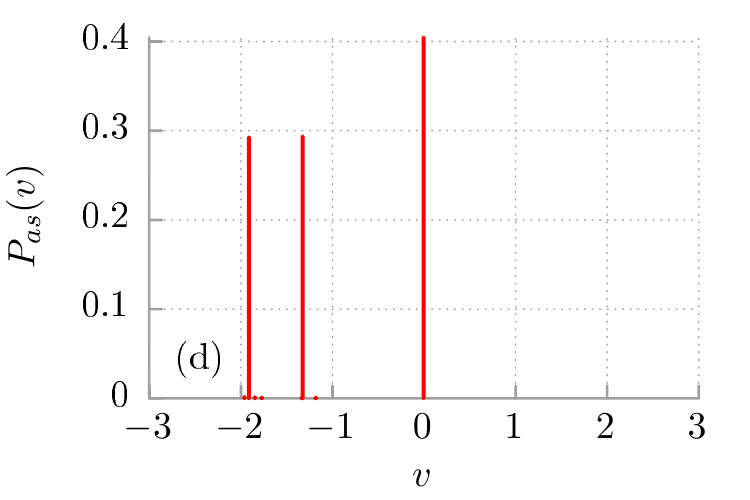}
\caption{Coexistence of absolute negative mobility and  ballistic diffusion. We refer the reader to Fig. \ref{fig1} for the description of quantities shown in the corresponding panels.
Parameters are: $a = 4.1$, $\omega = 1.618$,  $\gamma = 0.1445$ and $Q = 10^{-5}$. In panels (b)-(d) $f = 0.05$ for which $\boldsymbol{\alpha} = 2.006$ in panel (a). The fundamental period of driving force equals $\mathsf{T} = 3.88$.}
\label{fig2}
\end{figure}

\subsubsection{Ballistic diffusion}

Yet another regime of AD which may coexist with ANM in the asymptotic long time limit relates to ballistic diffusion \cite{zaburdaev2015}. In this case the coordinate variance grows in time as $\langle \Delta x^2(t) \rangle \sim t^{\alpha}$  with  power exponent $\alpha = 2$. This behaviour is characteristic, e.g. for low dimensional Hamiltonian systems which usually posses mixed phase space and AD is related to the stickiness close to its ballistic regions \cite{gluck1998,denisov2002}. In Fig. \ref{fig2} we report the parameter regime for which  ballistic diffusion and  negative mobility coexist.

In particular, in panel (a) we present the directed velocity $\langle \mathbf{v} \rangle$, Eq. (11), and the power exponent $\boldsymbol{\alpha}$, Eq. (21), as a function of the external bias $f$. The reader may notice the corresponding force interval for which negative mobility and ballistic diffusion emerge simultaneously. In the depicted bias window the Lyapunov exponent is zero meaning that in the presented parameter regime the underlying dynamics is non-chaotic. Panel (b) of the same figure shows the time evolution of the directed velocity $\langle \mathbf{v}(t) \rangle$, Eq. (13), the power exponent $\boldsymbol{\alpha}(t)$, Eq. (20), (both left axis) and the diffusion coefficient $D(t)$, Eq. (17), (right axis). After the transient period the velocity $\langle \mathbf{v}(t) \rangle$  as well as the  exponent $\boldsymbol{\alpha}(t)$ reach their stationary values. The former is negative $\langle \mathbf{v}(t) \rangle < 0$ while the latter $\boldsymbol{\alpha}(t) = 2$,  i.e., it corresponds to the mentioned ballistic diffusion.

In panel (c) we investigate the impact of thermal fluctuations on both considered transport characteristics, i.e., $\langle \mathbf{v} \rangle$, Eq. (11) and $\boldsymbol{\alpha}$, Eq. (21). Negative mobility is rooted in the deterministic and non-chaotic dynamics as this phenomenon is present even for the limit of vanishing thermal noise intensity $Q \to 0$. Similarly, ballistic diffusion is observed for low to moderate temperature regimes. The origin of ballistic diffusion is depicted in  panel (d) where we show the asymptotic period averaged probability distribution $P_{as}(v)$, Eq. (12), in the deterministic system with $Q = 0$. The presented parameter regime possesses three attractors which are still clearly visible for low to moderate temperature. The first two corresponds to the solution transporting in the negative direction $\langle \mathbf{v} \rangle < 0$ whereas the third describes locked motion $\langle \mathbf{v} \rangle = 0$ for which the Brownian particle dwells in one or several potential wells. Therefore, the ensemble averaged particle coordinate scales asymptotically as $\langle x(t) \rangle \sim \langle \mathbf{v} \rangle t$. As a consequence the position variance  is proportional to time in the second power $\langle \Delta x^2(t) \rangle \sim t^2$. We stress that this is due to  existence of the additional attractor describing the locked solution $\langle \mathbf{v} \rangle = 0$ as otherwise the motion would be eventually with a constant dispersion, c.f. Fig. \ref{fig1}. Another scenario which could potentially lead to appearance of ballistic diffusion is the coexistence between two counter-propagating attractors \cite{spiechowicz2016scirep,spiechowicz2015pre}.
\begin{figure}[t]
\centering
\includegraphics[width=0.45\linewidth]{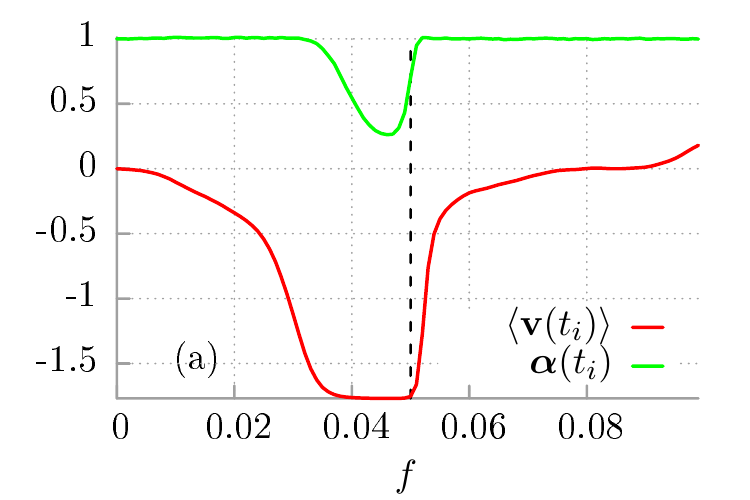}
\includegraphics[width=0.45\linewidth]{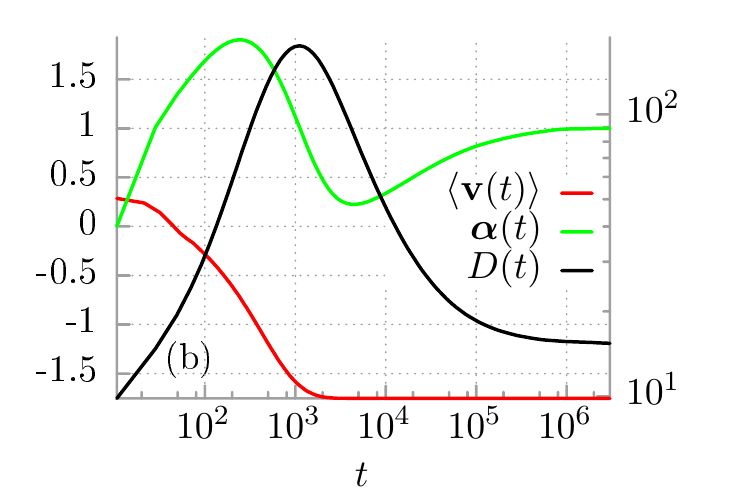}\\
\includegraphics[width=0.45\linewidth]{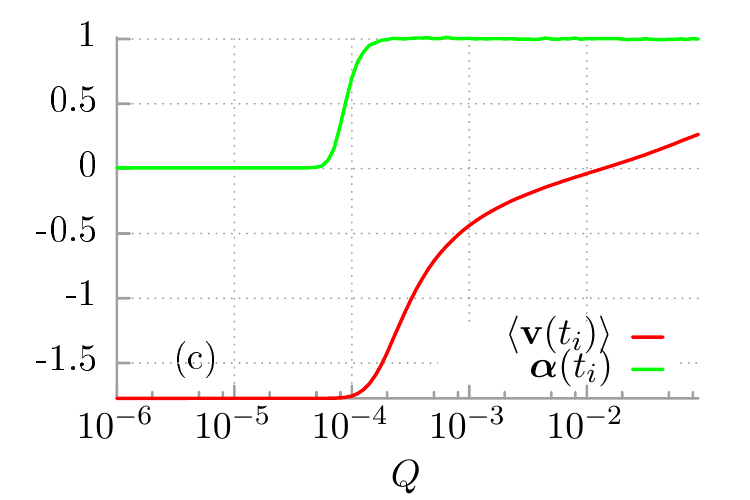}
\includegraphics[width=0.45\linewidth]{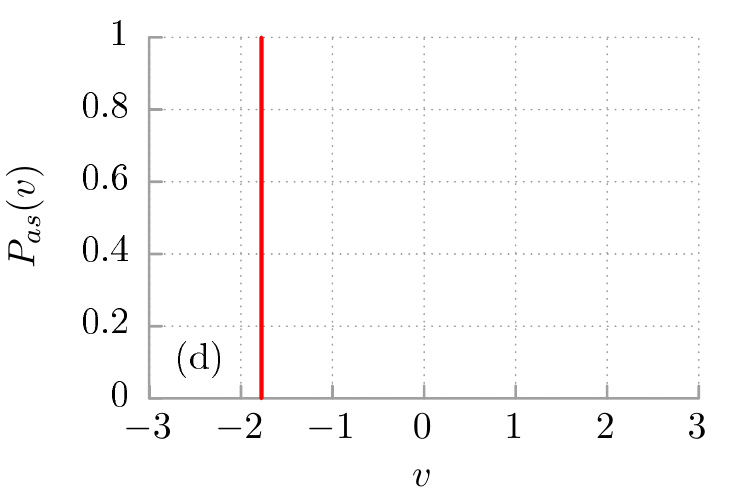}
\caption{Coexistence of absolute negative mobility and subdiffusion. We refer the reader to Fig. \ref{fig1} for the description of quantities shown in the corresponding panels. 
Parameters are: $a = 1.6$, $\omega = 0.59$, $\gamma = 0.126$ and $Q = 10^{-4}$. In panels (b)-(d) $f = 0.05$ for which $\boldsymbol{\alpha}(t_i) = 0.7$ with $t_i = 10^4 \,\mathsf{T}$ in panel (a). The fundamental period of driving force equals $\mathsf{T} = 10.64$.}
\label{fig3}
\end{figure}
\subsection{Regimes of transient anomalous diffusion}
\subsubsection{Subdiffusion}
The existence of subdiffusion is typically attributed to occurrence of broad probability densities of the residence times or strong correlations in the system dynamics. They can reflect physical properties of the setup like disorder, trapping, viscoelasticity of the medium or geometrical constraints \cite{metzler2014, bouchaud1990, meroz2015}. Since none of the above elements are present in our system we expect that we may observe subdiffusion only as a transient effect. We exemplify this scenario with our Fig. \ref{fig3}.

In panel (a) we present the directed velocity $\langle \mathbf{v}(t_i) \rangle$ and the power exponent $\boldsymbol{\alpha}(t_i)$ [both depicted for the intermediate elapsed time,  $t_i = 10^4 \,\mathsf{T}$], as a function of the external bias $f$. The reader will notice the bias window for which the directed velocity becomes negative and the power exponent assumes a value $\boldsymbol{\alpha} (t_i) < 1$. For this interval of bias values $f$  the Lyapunov exponent is equal to zero, meaning that the system is in a non-chaotic regime. 
In panel (b) of the same figure we depict the time evolution of the quantifiers of interest. In particular, the directed velocity $\langle \mathbf{v}(t) \rangle$ (left axis) reaches its stationary value  monotonically. In contrast, the power exponent $\boldsymbol{\alpha}(t)$ (left axis) and the diffusion coefficient $D(t)$ (right axis) display a non-monotonic behaviour with time. For increasing  $D(t)$ superdiffusion is observed ($\alpha > 1$) while for decreasing $D(t)$  subdiffusion occurs with $\alpha < 1$.
Overall the subdiffusive behaviour lasts three decades of the dimensionless time units and most likely it can be prolonged by proper tuning of the system parameters \cite{spiechowicz2016scirep,spiechowicz2017scirep}. It is important to note that the relaxation time of the directed velocity $\langle \mathbf{v}(t) \rangle$ and the power exponent $\boldsymbol{\alpha}(t)$ is radically different. The first takes about $10^3$ while the second needs $10^6$ dimensionless time units.

In panel (c) we investigate the impact  of thermal noise intensity, $Q \propto T$, on the directed velocity $\langle \mathbf{v}(t_i) \rangle$ and the power exponent $\boldsymbol{\alpha}(t_i)$ at time $t_i = 10^4 \,\mathsf{T}$. We observe that the origin of negative mobility lies in the deterministic and non-chaotic dynamics  since this phenomenon persists even in the limit of vanishing thermal noise intensity $Q \to 0$. On the contrary, subdiffusion is observed only for a narrow interval of temperatures. For smaller thermal noise intensity the motion is already with a constant dispersion while for larger the diffusion process is normal. The former fact combined with the observation that the considered parameter regime is non-chaotic allows to conclude that there must exist only one attractor transporting into  negative direction. In this respect this panel resembles the scenario depicted in Fig. \ref{fig1} (c). We confirmed this statement by investigation of the corresponding asymptotic period averaged probability distribution $P_{as}(v)$ which we depict in panel (d). An explanation of the mechanism standing behind the coexistence of negative mobility and subdiffusion in this regime  lies beyond the scope of this work. It would require a detailed analysis of the real time dynamics of the system including for example the possibility of the existence of ghost attractors in the phase space \cite{tel2011}. However, the panel (c) corroborates that surely thermal noise must play here an essential role \cite{spiechowicz2017scirep}.
\begin{figure}[t]
\centering
\includegraphics[width=0.45\linewidth]{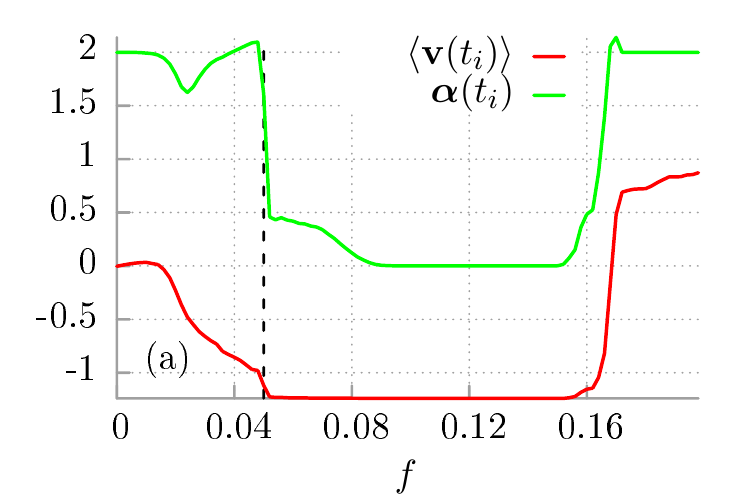}
\includegraphics[width=0.45\linewidth]{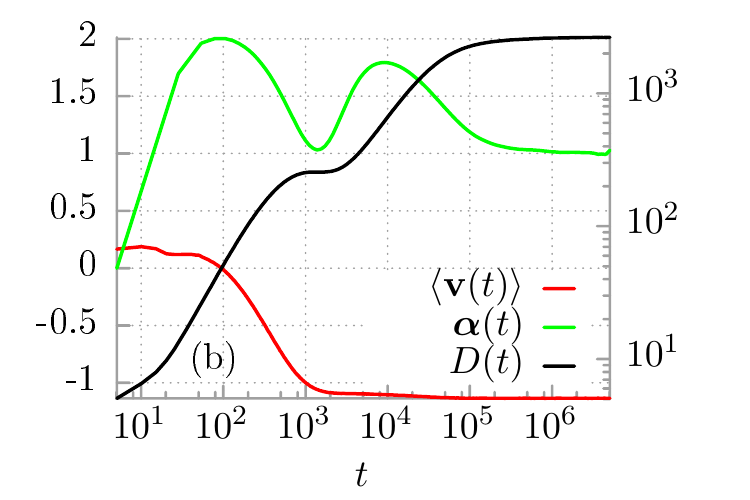}\\
\includegraphics[width=0.45\linewidth]{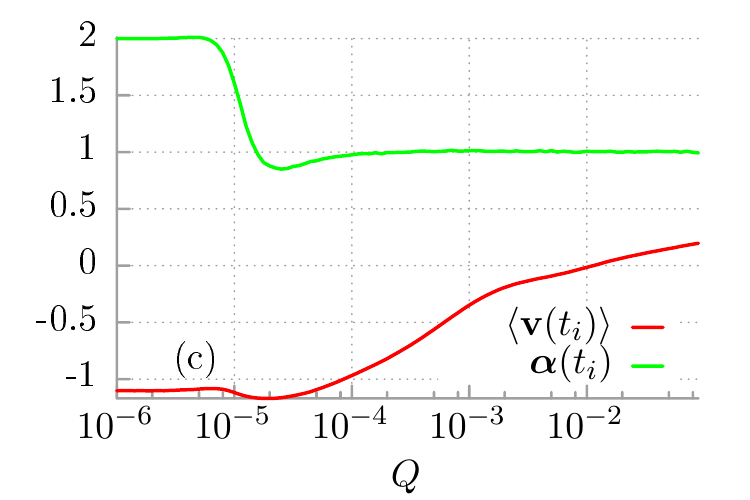}
\includegraphics[width=0.45\linewidth]{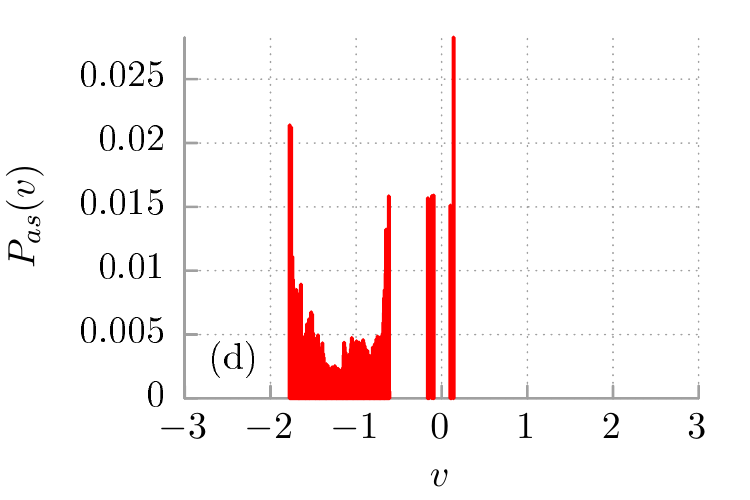}
\caption{Coexistence of  absolute negative mobility and superdiffusion. We refer the reader to Fig. \ref{fig1} for the description of quantities shown in the corresponding panels.
Parameters are: $a = 3.1$, $\omega = 1.24$, $\gamma = 0.21$ and $Q = 10^{-5}$. In panels (b)-(d) $f = 0.05$ for which $\boldsymbol{\alpha}(t_i) = 1.61$ with $t_i = 10^4 \,\mathsf{T}$ in panel (a). The fundamental period of driving force equals $\mathsf{T} = 5.06$.}
\label{fig4}
\end{figure}
\subsubsection{Superdiffusion} The emergence of superdiffusion is typically related to existence of broad probability densities of flights or strong correlations in the system. For instance, in chaotic dynamics of generic systems the origin of the long tails in the distribution of traps and flights relies on the stickiness of the chaotic trajectories close to the region of regular motion \cite{gluck1998,denisov2002}. This mechanism introduces long-term correlations which may lead to superdiffusion when stable ballistic trajectories exist, possibly resulting in Levy flights or Levy walks \cite{zaburdaev2015}. In Fig. \ref{fig4} we show the parameter regime within which both negative mobility and superdiffusion coexist.

In   panel (a) we depict the dependence of the directed velocity   and the power exponent (both depicted for the intermediate time scale $t_i = 10^4 \,\mathsf{T}$) versus the static bias. There is a wider interval of the force for $f<0.05$ and a narrower interval around $f \approx 0.165$ for which superdiffusion and negative mobility also concurrently exist.
Panel (b) of the same figure shows time evolution of the relevant quantifiers. In particular, we note that the detected superdiffusive anomaly, $1 < \boldsymbol{\alpha}(t) < 2$, merely presents  a transient effect and for $t\to\infty$ diffusion becomes normal. However, the lifetime of superdiffusion spans over $10^6$ dimensionless time units and is twice as long as the relaxation time of the driving-period averaged velocity $\langle \mathbf{v}(t) \rangle$. The reader may also draw attention to the non-monotonic behaviour of the time dependent power exponent $\boldsymbol{\alpha}(t)$.

In panel (c) we depict the influence of temperature $Q \propto T$ on negative mobility and superdiffusion at the time instant $t_i = 10^4 \,\mathsf{T}$. It is remarkable that a small dose of thermal noise enhances the negative response of the system while when being large - it destroys this effect completely. Depending on temperature and time scale different diffusive regimes are observed. In the limit of deterministic dynamics one encounters  ballistic diffusion with $\boldsymbol{\alpha}(t_i) = 2$. In the  low temperature regime both transient superdiffusion and subdiffusion emerge. In the high temperature regime,  diffusion is normal. Panel (d) depicts the probability density $P_{as}(v)$ for the deterministic counterpart of the system, i.e. with $Q = 0$. Contrary to the previously studied cases, in this figure one notices a dense regime of stable states which presents a characteristic signature for a rich chaotic dynamics. The complexity of the latter combined with the significant role of thermal noise (c.f. panel (c)) is reckoned to be most likely responsible for the detected transient superdiffusion as well as for negative mobility. However, we fail to provide reliable explanation for this complex behaviour, thereby leaving this challenging issue open for future research.
\begin{figure}[t]
\centering
\includegraphics[width=0.45\linewidth]{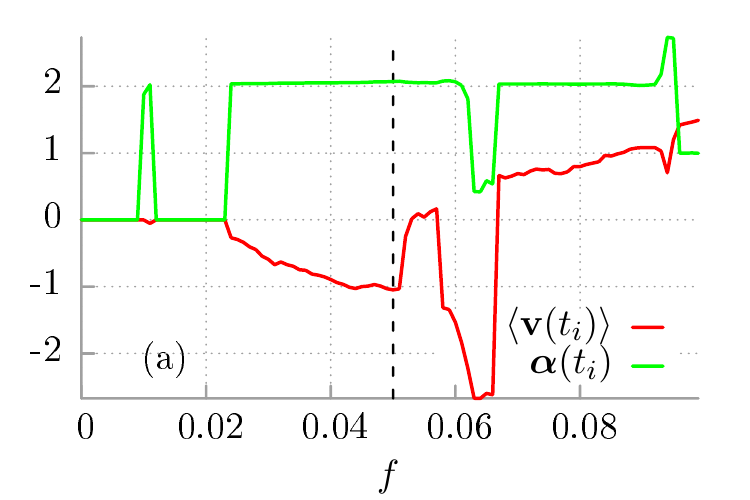}
\includegraphics[width=0.45\linewidth]{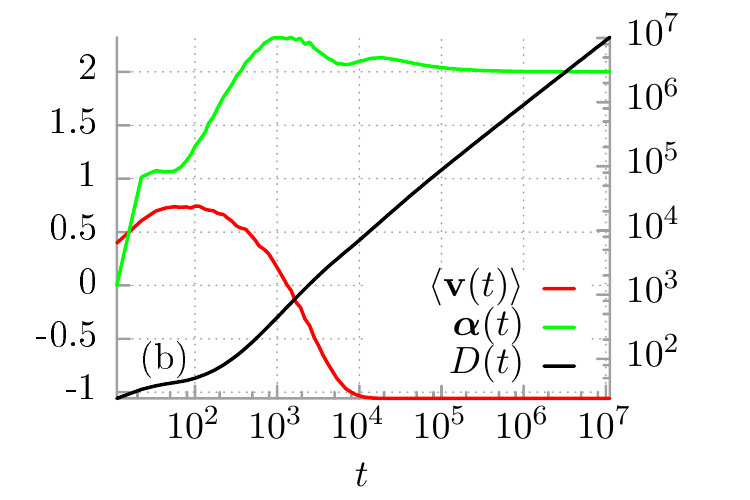}\\
\includegraphics[width=0.45\linewidth]{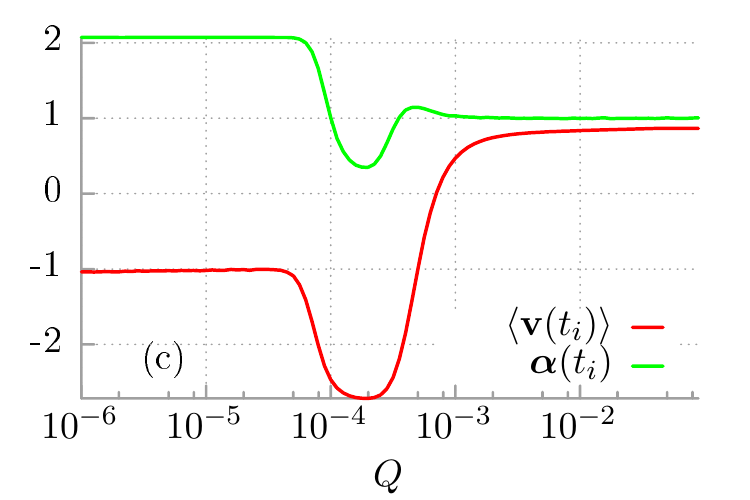}
\includegraphics[width=0.45\linewidth]{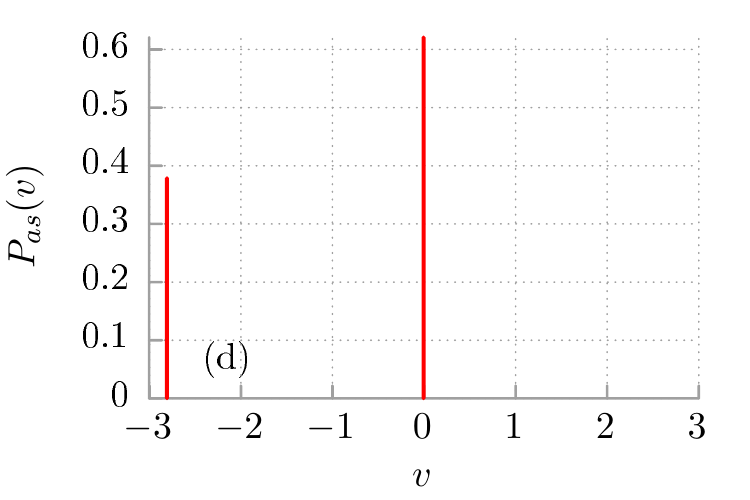}
\caption{Coexistence of absolute negative mobility and  hyperdiffusion. We refer the reader to Fig. \ref{fig1} for the description of quantities shown in the corresponding panels.
Parameters are: $a = 2.1$, $\omega = 0.56$, $\gamma = 0.072$ and $Q = 0$. In panels (b)-(d) $f = 0.05$ for which $\boldsymbol{\alpha}(t_i) = 2.07$ with $t_i = 10^4 \,\mathsf{T}$ in panel (a). The fundamental period of driving force equals $\mathsf{T} = 11.2$.}
\label{fig5}
\end{figure}
\subsubsection{Hyperdiffusion}
Finally, we discuss the possibility of the emergence for  hyperdiffusion (superballistic diffusion). This anomaly means that the coordinate variance grows faster than in the ballistic case, i.e.  $\alpha > 2$. An example of such behaviour was reported already in 1926 by Richardson who analysed the relative diffusion of two tracer particles in a turbulent flow and observed anomalous diffusion with characteristic cubic scaling of the mean squared displacement $\langle \Delta x^2(t) \rangle \sim t^3$ \cite{richardson1926}. Transient hyperdiffusion was detected also for generalized Brownian motion in tilted washboard potential and was attributed to transient heating of particles from thermal bath, c.f. Refs.  \cite{lu2007,siegle2010,zhang2017}. In Fig. \ref{fig5} we present here the parameter regime for which negative mobility and  such transient hyperdiffusion in fact  does coexist.

Panel (a) shows the  directed velocity $\langle \mathbf{v}(t_i) \rangle$ and the power exponent $\boldsymbol{\alpha}(t_i)$ (both depicted for the intermediate time scale $t_i = 10^4 \,\mathsf{T}$) as a function of the external bias $f$. Notably, in the vicinity of $f = 0.05$ the reader can detect the coexistence of hyperdiffusion with $\boldsymbol{\alpha}(t_i) > 2$ and  negative mobility $\langle \mathbf{v}(t_i) \rangle < 0$. In this interval of $f$  the Lyapunov exponent $\lambda$ is zero thus excluding the possibility of emergence of chaotic dynamics. This observation is confirmed in panel (b) where the time evolution of the corresponding observables is presented. In particular, we note that the asymptotic diffusive behaviour for this parameter regime is ballistic with $\boldsymbol{\alpha}(t) = 2$. However, transient hyperdiffusion lasts over  several decades of time.

Panel (c) illustrates the influence of thermal fluctuations on both, the directed  velocity and  diffusive behaviour expressed by the power exponent for the intermediate time at $t_i = 10^4 \,\mathsf{T}$. As the negative mobility persists even in the limit of deterministic dynamics $Q \to 0$ and the corresponding Lyapunov exponent is zero $\lambda = 0$ we conclude that this effect is rooted in the deterministic non-chaotic dynamics. Ballistic diffusion occurs in the asymptotic long time limit for low-to-moderate temperature regimes.
The latter can be observed up to a thermal noise intensity $Q = 10^{-4}$, c.f. panel (c). Moreover, in dependence on  temperature a whole range of anomalous diffusion processes can be detected at  intermediate times  as it is illustrated in panel (c). In panel (d) we depict the probability density $P_{as}(v)$ for the deterministic system. There occur two attractors, one corresponding to the running state transporting in negative direction, $\langle \mathbf{v} \rangle < 0$, as well as the remaining locked state, obeying $\langle \mathbf{v} \rangle = 0$ for which the particle dwells in one or several potential wells. As we discussed earlier such a structure of attractors is responsible for asymptotic ballistic diffusion. However, the mechanism of emergence of transient hyperdiffusion remains as of now not understood, thus requiring future research activities towards its final resolution.

\section{Conclusions}
With this work we answer the question whether it is possible to observe a coexistence of two anomalous transport manifestations, namely absolute negative mobility and anomalous diffusion of a nonlinear nonequilibrium inertial Brownian motion assisted by thermal fluctuations, i.e. a system in contact with heat bath of temperature $T$. In this context we merged two research fields that so far have been explored separately.

In doing so we explicitly considered  inertial Brownian particle moving in a periodic symmetric potential which in addition is exposed to a cosinusoidal external driving and additionally  driven by a constant bias. We reveal parameter regimes for which a rich spectrum of anomalous diffusion processes and the negative mobility phenomena coexist. The former includes transport with a constant dispersion, subdiffusion, superdiffusion, ballistic diffusion and even  hyperdiffusion. Subdiffusion, superdiffusion and hyperdiffusion are only the transient effects in this setup.  Their corresponding persistence can last, however, over many orders longer than the intrinsic characteristic time scales of the system. We demonstrate that thermal fluctuations play important role in the emergence of anomalous diffusion processes. For a given parameter set depending on temperature  different anomalies may arise. It is important to note that in almost all presented cases the negative mobility phenomenon is rooted in the deterministic non-chaotic dynamics. We find that the coexistence of both anomalous diffusion and negative mobility is scarce and requires sensitive detailed tuning to appropriate parameter regimes in the underlying five-dimensional parameter space. It makes them very exceptional.

Our research may contribute to further understanding of peculiar transport phenomena appearing in microworld. The appealing strength and beauty of Brownian motion with its intrinsic Gaussian fluctuations lies in its universality and therefore our findings can be straightforwardly tested experimentally with a wealth of physical systems \cite{risken1996,thomas,kautz1996,blackburn2016,spiechowicz2014prb,renzoni,fulde1975,dieterich1980}. Most promising ones in this regard involve driven optical lattices and Josephson junctions.

\section*{Acknowledgement}
This work was supported by the Grants No. NCN 2017/26/D/ST2/00543 (J.S.) and No. NCN 2015/19/B/ST2/02856 (J.{\L}.)

\section*{Appendix}
The system analysed in this paper comprises a rich and complex five-dimensional parameter space $\{\gamma, a, \omega, f, Q\}$. Nevertheless, we performed numerical analysis with unprecedented resolution. We considered over $10^6$ different parameter sets. The high precision was made possible due to the innovative computational method which is based on employing GPU supercomputers, for details see Ref. \cite{spiechowicz2015cpc}. We employed a weak 2nd order order predictor-corrector method \cite{platen} to simulate the stochastic dynamics given by (\ref{dimless-model}). Since we are interested not only in a short time behaviour of the system, but also its asymptotic state, numerical stability is an extremely important problem to obtain reliable results. Hopefully, predictor-corrector algorithm is similar to implicit methods but does not require the solution of an algebraic equation at each step. It offers good numerical stability which it inherits from the implicit counterpart of its corrector.

Since we were dealing with different parameter regimes we integrated Eq. (\ref{dimless-model}) by the predictor-corrector method with the time step scaled by the fundamental period $\mathsf{T} = 2\pi/\omega$ of the dynamics, i.e. $h = 10^{-2} \times \mathsf{T}$. The initial positions $x(0)$ and velocities $v(0)$ were uniformly distributed over the intervals $[0, 2\pi]$ and $[-2,2]$, respectively. Each time we checked that the corresponding results are not affected by a different choice of initial setting. The latter modifies only the transient dynamics of the setup which typically lasts only for a decade or two, see the section before Discussion in Ref. \cite{spiechowicz2016scirep}. It does not alter neither the intermediate nor the asymptotic behaviour of the system. The quantities of interest were averaged over the ensemble of $2^{17} = 131072$ trajectories, each starting with different initial condition according to the distributions presented above. The number of realizations of stochastic dynamics is not accidental and was chosen carefully to maximize the performance of the numerical simulation, see Ref. \cite{spiechowicz2015cpc} for details. The quantities of interest were calculated after each fundamental period $\mathsf{T} = 2\pi/\omega$ in the time interval $[\mathsf{T}, 10^6 \mathsf{T}]$ in which either several transport quantifiers averaged over the driving period or their asymptotic stationary values  are evaluated in tailored parameter regimes. This also allowed to classify those pertinent quantifiers as transient only or reaching also a unique stationary asymptotic long time value.

\end{document}